# Surface State Transport and Ambipolar Electric Field Effect in $Bi_2Se_3$ Nanodevices

Hadar Steinberg[1], Dillon R. Gardner[1], Young S. Lee[1], Pablo Jarillo-Herrero[1,*]

1. Department of Condensed Matter Physics, Massachusetts Institute of Technology, 77 Massachusetts Ave, Cambridge MA 02139, USA

* pjarillo@mit.edu

**Electronic transport experiments involving the topologically protected states found at the surface of $Bi_2Se_3$ and other topological insulators require fine control over carrier density, which is challenging with existing bulk-doped material. Here we report on electronic transport measurements on thin (<100 nm) $Bi_2Se_3$ devices and show that the density of the surface states can be modulated via the electric field effect by using a top-gate with a high-*k* dielectric insulator. The conductance dependence on geometry, gate voltage, and temperature all indicate that transport is governed by parallel surface and bulk contributions. Moreover, the conductance dependence on top-gate voltage is ambipolar, consistent with tuning between electrons and hole carriers at the surface.**

Topological insulators (TIs)[1-4] constitute a new class of materials with unique properties resulting from the relativistic-like character and topological protection of their surface states[5,6]. Theory predicts these to exhibit a rich variety of physical phenomena such as anomalous magneto-electric coupling[7] and Majorana excitations[8]. Although TI surface states have been detected in Bi-based compounds by ARPES[9-11] and STM techniques[12-14], electrical control over their density, required for most transport experiments, remains a challenge. Existing materials are heavily doped in the bulk, thus preventing electrical tunability of the surface states and their integration into topological quantum electronic devices.

$Bi_2Se_3$ is a suitable platform to demonstrate electronic transport physics through topologically protected surface states due to its relatively wide bulk band-gap (0.3 eV)[9]. In bulk $Bi_2Se_3$ an anomalous high field magnetoconductance was reported[15], and an indication of surface transport was found in the form of Shubnikov-de-Haas oscillations in $Bi_2Se_3$[16] and $Bi_2Te_3$[17]. In thin layers of doped $Bi_2Se_3$, a significant part of the conductance should take place through the top and bottom surface states as well as through the bulk. Electronic transport was studied on $Bi_2Se_3$ nano-platelets[18] and nanoribbons[18,19], where Aharonov-Bohm interference[19] was interpreted as coherent surface transport around the ribbon[19].

In this letter we report on transport measurements of exfoliated $Bi_2Se_3$ nanodevices of variable thickness, establishing the different contributions of bulk and surface states to the device conductance. Moreover, by using top and bottom gate electrodes, we are able to modulate the conductance via the electric field effect, including an ambipolar regime for the top surface state, consistent with the gapless band structure of the TI surface states. Our measurements enable us to estimate the densities and mobilities of the bulk and the surface. We measure the temperature dependence of the device conductance, and extract the evolution of the surface mobility with temperature, which allows to identify possible scattering mechanisms for the surface states.

Arsenic-doped $Bi_2Se_3$ single crystals are synthesized by melting a stochiometric mixture of Bi and Se, and trace amounts of As, in a quartz tube at 850°C, followed by slow cool down. Infrared reflectometry and electronic resistivity measurements indicate that the material is electron-doped, with resistivity ranging from 0.4 to 0.7 mΩcm at room temperature. The resulting ingots cleave easily, and are exfoliated to produce thin flakes, which are deposited on a Si substrate capped with $SiO_2$ and contacted using standard electron beam lithography. An AFM image of a device contacted in a Hall bar geometry is shown in Figure 1a. The device scheme is presented in Figure 1c: The two-gates are formed by using the doped Si as a back gate electrode with the $SiO_2$ layer as the dielectric, and a lithographically defined metallic contact (Ti/Au), as a top-gate, with a high-$k$ dielectric layer (20 nm thick $HfO_2$ or $Al_2O_3$) deposited by ALD (Atomic Layer Deposition). We have measured in total 20 devices, all exhibiting qualitatively similar behavior.

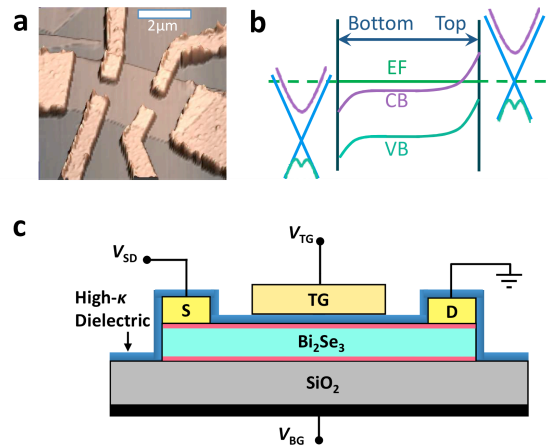

**Figure 1. Device geometry. a,** AFM image of a 17nm thick $Bi_2Se_3$ device contacted in a Hall bar geometry. **b**, Schematic variation of the band structure along the $z$ direction, showing the bulk conduction (purple) and valence (green) bands bending. The topological surface states bands (light blue) can be shifted

by application of a gate voltage or surface doping. **c,** Scheme of a back-gated / top-gated device: The device appearing in **a** (light blue) is coated by a layer of high-*k* dielectric (dark blue), followed by evaporation of a Ti/Au gate. The device is also gated from the bottom, by a doped-Si back-gate. The top and bottom surface states are indicated by red lines.

We demonstrate the contribution of the surface channels to conduction by measuring the square conductance *G* and the Hall coefficient $R_H$, on devices of varying thickness *d*. *G* includes contributions from the bulk and surface carriers:

$$G = \sigma_{bulk} d + \sigma_{surface} \quad (1)$$

where *σ* is the conductivity associated with each component. Specifically $\sigma_{bulk} = n_{bulk}\mu_{bulk}e$ and $\sigma_{surface} = n_{surface}\mu_{surface}e$, where $n_{bulk,surface}$ are the bulk and surface densities, $\mu_{bulk,surface}$ are their mobilities and *e* the absolute value of the electron charge. Figure 2a shows *G* vs *d* for a set of devices fabricated from the same ingot (denoted Ingot A). The data agree well with the linear dependence predicted by Equation 1 (solid line), confirming that all devices have similar bulk and surface properties. From the intercept at *d* = 0 we can estimate $\sigma_{surface} \sim 200\ e^2/h$, suggesting that a considerable fraction of the total current is associated with the surfaces.

Additional information can be obtained by examining the Hall coefficient $R_H$ for the same set of devices (Figs. 2b-c). We limit the discussion to magnetoconductance at low magnetic fields *B* since in $Bi_2Se_3$ $R_H$ could exhibit non trivial effects at high magnetic fields[20]. In conductors with a single type of charge carrier $R_H = -1/ne$, where *n* is the carrier density. $R_H$ has different units for 3D and 2D conductors, due to the different dimensionality of *n*. Therefore, we distinguish between $R_H^{3D}$ and $R_H^{2D}$ depending on whether *n* is a volume or surface density. $R_H^{3D}$ is extracted from the Hall voltage $V_H = -IB/ned$ by taking $R_H^{3D} = V_H d/BI$, which for a pure 3D conductor is equal to $-1/ne$, i.e., independent of $d$. $R_H^{2D}$ is also extracted from the Hall voltage, which for a 2D system or conducting surface is $V_H = -IB/ne$, by taking $R_H^{2D} = V_H/BI$. $R_H^{2D} = -1/ne$ for a pure 2D system. Therefore, if a system exhibits pure 3D bulk or pure 2D surface transport, the corresponding $R_H$ (properly normalized) is independent of device thickness. Figure 2b shows the measured $R_H^{3D}$ versus $d$, demonstrating that $R_H^{3D}$ clearly depends on $d$, confirming that conductance deviates from a simple bulk 3D model. To test if the current is carried exclusively by surface states we extract $R_H^{2D}$ from the same $V_H$ dataset (Figure 2c). $R_H^{2D}$ also depends on $d$, indicating that the electronic transport is not purely two-dimensional.

If the system conductance is neither purely 3D nor 2D, we have to consider the parallel contribution of both bulk and surface. This is done by deriving a two population charge carrier model, similar to those used for semiconductor heterostructures, where multiple carriers of different mobilities, such as electrons and holes, or electrons belonging to different bands, contribute to conductance in parallel[21]. In the two-carrier model $R_H$ depends on the respective densities of the two carriers and the ratio of their mobilities (see SI for details). In our case, one carrier is a surface channel and the other is a bulk channel, setting $\alpha = \mu_{bulk} / \mu_{surface}$ we have

$$R_H^{2D} = -\frac{\left(n_{bulk}\alpha^2 d + n_{surface}\right)}{e\left(n_{bulk}\alpha d + n_{surface}\right)^2}; \quad R_H^{3D} = R_H^{2D} d \qquad (2)$$

Although the available dataset spans a limited range of device thicknesses (17-80 nm), the densities and mobilities can be estimated by fitting equation 2 (solid line in panels b,c) to the data. This yields $n_{surface} \sim 4 \cdot 10^{13}$ cm$^{-2}$ and $n_{bulk} \sim 1 \cdot 10^{19}$ cm$^{-3}$. The quality of the fit is relatively insensitive to the parameter $\alpha$ (with values from 0.5 to 2 yielding similarly good fits), which does not allow a precise determination of the mobility ratio using $R_H$. However we can use the conductance data: plugging $n_{surface}$ and $n_{bulk}$ into the slope and intercept found by fitting the data in panel (a) yields $\mu_{bulk} \sim 1700$ cm$^2$/Vs and $\mu_{surface} \sim 1000$ cm$^2$/Vs. Both $n_{bulk}$ and $\mu_{bulk}$ agree with independent measurements carried out on bulk samples of the same material.

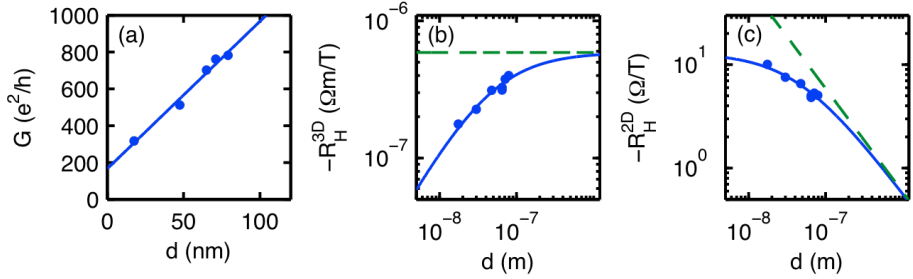

**Figure 2: Evidence for surface transport.** (a) Conductance $G$ vs. thickness $d$ for a set of 5 devices exfoliated from the same bulk ingot. The solid line is a fit to Equation 1. (b) Hall coefficient $-R_H$ (in 3D units) vs. $d$. The data are taken for the same 5 devices as in (a), plus 2 additional devices. The solid line is a fit to the two-carrier model (equation 2). The dashed green line represents a constant value, expected if only bulk carriers are present. (c) Same as (b) given in 2D units.

The above results demonstrate the presence of a surface conducting channel but does not reveal which of the two surfaces, top or bottom (or both) is involved in transport. We can differentiate the contributions of both surfaces by using the top and back gate electrodes to separately tune their densities. Figure 3a shows $R$ vs. top gate voltage $V_{TG}$. $R$ is the square resistance measured in a 4-probe geometry. We report data taken on two devices at $T = 4$K, device 1 (40 nm thick) and device 2 (45 nm). The devices are fabricated from an ingot denoted as Ingot E, (of similar bulk properties to Ingot A) with HfO$_2$ dielectric. Devices of both ingots reported in this study exhibit the same behavior. The gate induces a modulation of 1-2% over a resistance of 40-60 $\Omega$, with a resistance peak near $V_{TG} = 0$ V. In some devices (e.g. Device 1), a sharp drop in resistance is found at negative voltage, which we associate with the onset of bulk conduction via the bulk valence band states at the surface (see SI). To separate the parallel contributions of bulk and surfaces it is convenient to discuss square conductance $G = 1/R$, which decomposes into $G(V_{TG}) = G^{bottom}(V_{TG}) + G^{bulk}(V_{TG}) + G^{top}(V_{TG})$. The minimum conductance $G(V_{min})$ is of order a few hundred $e^2/h$, mostly associated with $G^{bottom} + G^{bulk}$. Figure 3c shows $G-G(V_{min})$, singling out the gate-induced modulation from the relatively large background. The conductance has ambipolar characteristics: $G(V_{TG} > V_{min})$ has a positive slope expected for electron conductance, and $G(V_{TG} < V_{min})$ has negative slope, characteristic of hole conductance. The electron to hole transition at $V_{min}$ is the expected behavior for an electron system with a Dirac dispersion[22] where $G = \sigma = |n|e\mu$ reaches a minimum value when $n = 0$. We note that due to the large background of bulk signal the minimum conductance of order $e^2/h$ expected for Dirac fermions[23] cannot be extracted. For the same reason, it is not possible to detect the transition from negative to positive carriers in the Hall coefficient (not shown).

In most of our samples the electron and hole slopes are asymmetrical. It is likely that this asymmetry is at least partly due to the bulk channel, since modulating the density of the surface state necessarily leads to changes of bulk charge as well (see SI) and hence $G(V_{TG})$ may be decomposed into a symmetrical component $G^S(V_{TG})$ associated mostly with the surface and a linear component associated mostly with the bulk.

One of our key findings is that the conduction minima in $HfO_2$ gated devices appears near $V_{TG} = 0$, suggesting that although the bulk is highly doped, the surface band structure is shifted so that the charge neutrality point of the surface state is close to the Fermi energy, as shown in Figure 1b. This result is found in more than 10 devices fabricated in separate batches, and suggests that it is possible to tune the surface state Fermi energy by careful design of the interface. This shift is likely associated with the nature of the $Bi_2Se_3$ – $HfO_2$ interface, where metal-induced gap states are expected to form[24]. Due to its large dielectric constant, $HfO_2$ can cause a significant shift in the work function of metals[25,26], bending the bands upwards, and leaving a layer depleted of bulk carriers near the surface. To further test the effect of the dielectric on the top surface band structure, we have fabricated devices with $Al_2O_3$, which has a smaller dielectric constant, and found the conductance minimum is shifted to $V_{TG} \sim$ -10V (devices 3 and 4, Figure 3b).

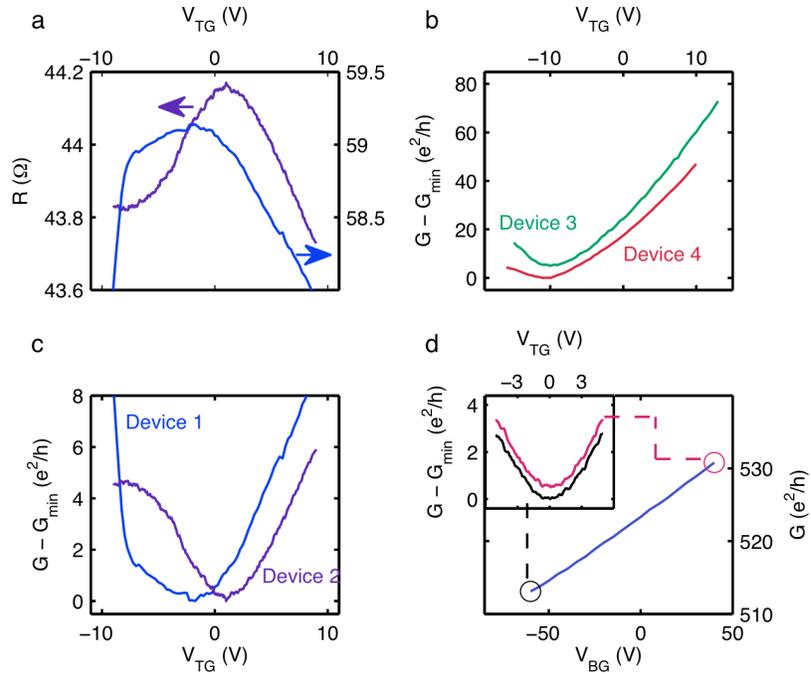

**Figure 3: Electric field effect measurements. a,** Device resistance $R$ vs. top gate voltage $V_{TG}$ for devices 1 and 2, of thickness 40 nm and 45 nm respectively. $R$ exhibits ~2% modulation, with a peak near $V_{TG} \sim$ 0V. **b,** $G$-$G(V_{min})$ vs. $V_{TG}$ for devices 3 and 4 ($Al_2O_3$ dielectric). **c,** $G$ - $G(V_{min})$ vs. $V_{TG}$ for devices 1 and 2 ($HfO_2$ dielectric). A parallel conductance $G(V_{min})$ of 436 $e^2/h$ and 583 $e^2/h$, respectively, is subtracted. **d,** $G$ vs. back gate voltage $V_{BG}$ for Device 2. Inset: $G$ - $G(V_{min})$ vs. $V_{TG}$ scans taken at $V_{BG}$ = +40

V (black) and $V_{BG}$ = -60 V (red), vertically shifted for clarity. The scans perfectly overlap, indicting that the top and bottom surfaces are gated independently.

The top gate measurements presented in Figure 3 point to a significant difference between the top and bottom surfaces: The top surface contributes 10-20 $e^2/h$ to the total conductance in all devices. This indicates that the surface conductance of ~200 $e^2/h$ found by fitting the data in Figure 2a is predominantly carried by the bottom surface. The top surface of the $Bi_2Se_3$ strongly couples to the high-$k$ dielectric, whereas the bottom surface only couples loosely to the underlying $SiO_2$. ARPES studies indicate that at the cleaved surface the $Bi_2Se_3$ energy bands bend downwards. This is consistent with the estimated density of the bottom surface, $n_{surface}$ ~ $4\times10^{13}$ cm$^{-2}$, which corresponds to a bottom surface Fermi energy $E_F$ ~ 0.6 eV with respect to the Dirac point (Figure 1b). We also note that the conductance is carried by surface states and not by the accumulated bulk carriers near the surface: The added contribution of the accumulation layer can be estimated as $\Delta G_{accumulation}$ ~ $W(n_{accumulation}-n_{bulk})\mu_{bulk}e$, $W$ being the thickness of the accumulation layer, which is in the range of few nm. This yields ~ 10-20 $e^2/h$, an order of magnitude less than the observed surface conductance.

Biasing the back-gate (Figure 3d) modulates the conductance, but a minimum feature is not detected, which is not surprising in view of the high carrier density of this surface. The field effect mobility is usually extracted by taking $e\mu = \partial\sigma/\partial n$, $en = CV_{BG,TG}$ where $C$ is the gate capacitance per unit area. If the entire change can be associated with the surface, than $\Delta G = \Delta\sigma$. However the gate charges both the bulk and the surface states, and a detailed model including screening and band bending, which is beyond the scope of this work, is required to account for the exact amount of charge induced on each channel.

Assuming that all the charge is induced on the bottom surface state leads to a lower bound for its mobility, $\mu \geq 600$ cm$^2$/Vs, consistent with the value obtained above. We note that varying the top gate voltage at different back gate voltages (inset to Figure 3d) confirms that the top and bottom channels are independent. It is also not trivial to extract the top surface mobility, due to the combined charging of the bulk and surface (screening due to the surface states in this case is weaker because of the proximity to the Dirac point). We find therefore a lower limit of $\mu \geq 50$ -100 cm$^2$/Vs, although it is most likely significantly higher.

We now turn to the temperature dependence of the conductance. Figure 4a shows $R(V_{TG})$ for selected temperatures $T$. A number of features evolve with $T$: (*i*) The background resistance increases, apparent as a vertical shift in $R(V_{TG})$. In the figure, the resistance plots taken at elevated temperatures are vertically shifted down to fit in the same plot (the actual bulk + bottom surface resistance $R(V_{min},T)$ is shown in the inset to panel d, and follows a similar dependence as that observed for bulk Bi$_2$Se$_3$[27]). (*ii*) The resistance peak flattens as $T$ is increased, accompanied by (*iii*) an increase of the background slope. This data lends more support for associating the symmetrical component $G^S(V_{TG})$ with the surface conductance: $G^S(V_{TG})$ is plotted in panel b for a range of temperatures, and appears to grow smoothly shallower as $T$ increases, consistent with a reduction in carrier mobility. In Figure 4c the full data set, at all temperatures studied, is presented as a color map.

We trace the relative change in the field effect mobility $\mu_{top}(T)$, shown in panel d, by taking the maximum derivative of $G^S(V_{TG})$ at each $T$. $\mu_{top}(T)$ remains almost unchanged up to $T = 50$K, and scales as $T^{-1.4}$ for $T > 50$K. Such power-laws are found ubiquitously in electronic transport and are typically a consequence of phonon scattering. The two most likely candidates in this case are $Bi_2Se_3$ acoustic phonons, which yield a similar slope in bulk $Bi_2Se_3$[27], or polar surface phonons in the $HfO_2$, which have been shown to strongly suppress the mobility in graphene devices[28]. At $T < 50K$ the mobility saturates, indicating scattering by static impurities. The linear dependence of $G^S(V_{TG})$ is suggestive of similar scattering mechanisms as those observed in graphene devices (including, for example, charged impurities[29,30]).

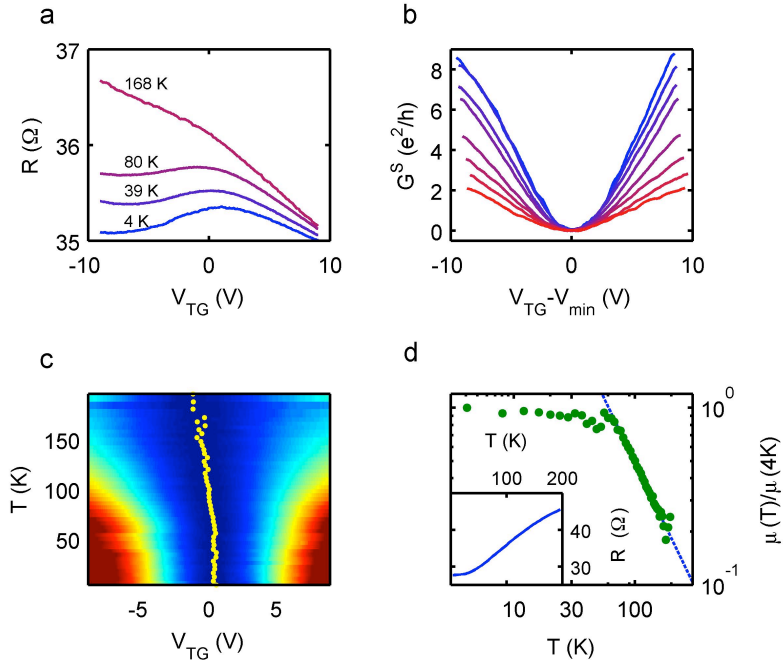

**Figure 4: Temperature dependence of the field effect mobility. a,** $R$ vs. $V_{TG}$ taken for Device 2 (45 nm thick) at $T = 4.1K$, $39.2K$, $80.1K$ and $167.8K$. Except for the lowest $T$, all data are shifted vertically for clarity. **b,** $G^S$ plotted vs. $(V_{TG} - V_{min})$, e.g. shifting the minimum conductance to 0. The data are obtained from a scan similar to **a**, taken for device 5 (55nm thick), by inverting and symmetrizing. **c,** Color map of $G^S(V_{TG})$ for device 5 (see (b) for vertical scale), red stands for high conductance, blue for low. The minimum conductance voltage $V_{min}$ is marked by a yellow dot for each $T$. **d,** Relative change in field effect mobility $\mu$ at the $V_{TG} > V_{min}$ branch, extracted from **c**. The mobility scales as $T^{-1.4}$ for $T > 50K$. Inset: Background resistance $R(V_{min})$ vs. $T$ (device 5), associated with bulk and bottom surface contributions.

In summary, we show that the surface states contribute significantly to the conductance of nanoscale $Bi_2Se_3$ devices and their contribution can be tuned via the electric field effect, even at relatively high bulk doping. The coexistence of three parallel conductance channels is secured by the topological protection of the surface states, opening new possibilities for electronic devices, scalable down to a thickness of few nanometers[31] while retaining the metallic nature of the surface states.

# Surface State Transport and Ambipolar Electric Field Effect in $Bi_2Se_3$ Nanodevices

Hadar Steinberg, Dillon R. Gardner, Young S. Lee, Pablo Jarillo-Herrero

**Supporting Information**

*Two-carrier surface-bulk model*

To calculate the Hall voltage in magnetic field $B$ in a general multiple-carrier model, one sums over the contributions of all carriers to the conductivity tensor $\sigma^{tot}$:

$$\sigma_{xx}^{tot} = \sum_i \frac{n_i e \mu_i}{(1+\mu_i^2 B^2)}; \quad \sigma_{xy}^{tot} = \sum_i \frac{n_i e \mu_i^2 B}{(1+\mu_i^2 B^2)}$$

Where the sum is over the carrier index $i$, $n_i$ and $\mu_i$ stand for density and mobility of carrier $i$, respectively, and $e$ is the absolute value of the electron charge. $\rho_{xy}$ is the off diagonal element of $(\sigma^{tot})^{-1}$. For two carriers:

$$\rho_{xy}(B) = -\frac{B}{e} \frac{(n_1\mu_1^2 + n_2\mu_2^2) + B^2 \mu_1^2 \mu_2^2 (n_1+n_2)}{(n_1\mu_1 + n_2\mu_2)^2 + B^2 \mu_1^2 \mu_2^2 (n_1+n_2)^2}$$

and for small $B$:

$$\rho_{xy}(B) = -\frac{B}{e} \frac{(n_1\mu_1^2 + n_2\mu_2^2)}{(n_1\mu_1 + n_2\mu_2)^2}.$$

It is common to use the Hall coefficient $R_H$:

$$R_H = \frac{\rho_{xy}}{B} = -\frac{1}{e} \frac{(n_1\mu_1^2 + n_2\mu_2^2)}{(n_1\mu_1 + n_2\mu_2)^2}.$$

When expressed in 3D units, $n$ being a volume density, we substitute $n_1 = n_{bulk}$, $\mu_1 = \mu_{bulk}$, $n_2 = n_{surface}/d$, $\mu_2 = \mu_{surface}$:

$$R_H = -\frac{1}{e} \frac{(n_{bulk}\mu_{bulk}^2 + n_{surface}\mu_{surface}^2/d)}{(n_{bulk}\mu_{bulk} + n_{surface}\mu_{surface}/d)^2}.$$

Setting $\alpha = \mu_{bulk} / \mu_{surface}$ we finally arrive at Eq. 1 of the main text describing the dependence of $R_H^{2D}$ and $R_H^{3D}$ on $d$:

$$R_H^{2D} = -\frac{\left(n_{bulk}\alpha^2 d + n_{surface}\right)}{e\left(n_{bulk}\alpha d + n_{surface}\right)^2}; \quad R_H^{3D} = R_H^{2D} d$$

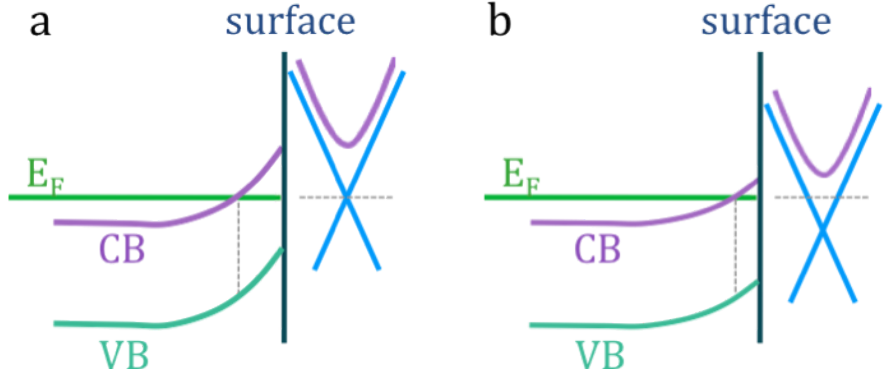

**Figure S1: Sinultaneous charging of the bulk and surface states. (a) the dispersion of the bulk conduction band (purple) and surface states (blue), is drawn on the right. The position dependence of the bulk bands is plotted on the left. The surface state is tuned such that the Dirac point is at the Fermi energy. The bottom of the bulk band is at 0.2eV. The bulk bands bend near the surface, forming a depletion layer (vertical dashed line) (b) As a gate voltage is applied and charge is added the dispersions of both bulk and surface states are shifted downward. The change in the bulk dispersion results in a narrower depletion layer.**

*Simultaneous charging of bulk and surface*

We noted in the main text that based on geometry considerations, presented in Figure 2, both bulk and surface channels contribute significantly to the electronic transport. Here we argue that such bulk contribution should be detectable also in gating measurements. We demonstrate that the surface and bulk have to charge together using the energy band diagram in Figure S1. When the surface denstiy is zero the Dirac point is at the Fermi energy (a). Since the bulk conduction band is 0.2eV above the surface Dirac point, it will have to be above the Fermi energy, and hence undergo band-bending from bulk to surface, resulting in a depletion layer. The thickness of the depletion layer could be in the order of 1nm for the bulk densities in our samples. Charging the surface states results in a downward shift of their dispersion with respect to the Fermi energy (b). The bulk dispersion shifts downwards at the surface and within the depletion layer, narrowing of the depletion layer and hence charging the bulk. The charge added by the applied gate voltage is therefore shared between the surface states and the bulk. Note that this is not a consequence of screening, but rather of the coupling of surface and bulk dispersions which reside in the same band-structure. Screening plays an additional role in determining the amount of charge induced at the surface versus the bulk.

*Compilation of ΔG($V_{TG}$) Results*

If the application of gate voltage modifies the charge of the bulk, this should have a signature in the field effect measurements presented in Fig 3. We can notice this effect in Figure S2, where we plot $\Delta G(V_{TG}) = G(V_{TG})-G(V_{min})$ measured for 7 different devices with $HfO_2$ dielectric, fabricated on the same sample from Ingot E. This data set included the devices discussed in the main text (Devices 1, 2 and 5 marked on the panels).

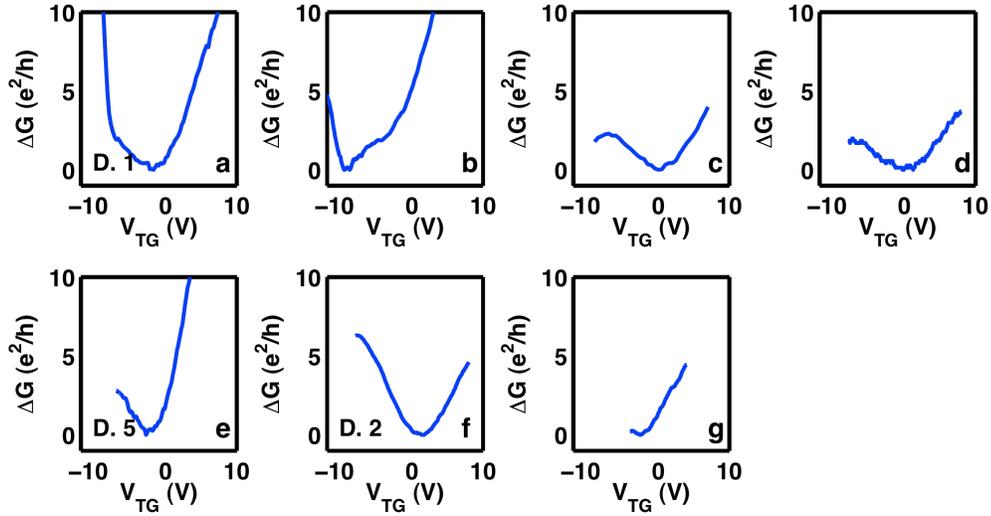

**Figure S2: Collection of ΔG($V_{TG}$) of 7 devices fabricated from Ingot E. Devices 1, 2 and 5 discussed in the main text are marked.**

If we symmetrize the data by subtracting a linear component from each data set (Figure S3) we find all the data sets are similar, with small variability of the minimal conductance feature around $V_{TG} \sim 0V$, and a typical slope $\Delta G/\Delta V_{TG}$. 2 out of the 7 devices exhibit a sharp increase in conductance at $V_{TG} \sim -8V$.

The conductance of each device can therefore be modeled as a sum of three contributions, assiciated with the bulk and both surfaces, where both the bulk and the top surface which on the gate voltage:

$G^{tot} = G^{bot} + G^{bulk}(V_{TG}) + G^{top}(V_{TG})$, where $G^{bulk}(V_{TG})$ is linear and $G^{top}(V_{TG})$ is ambipolar.

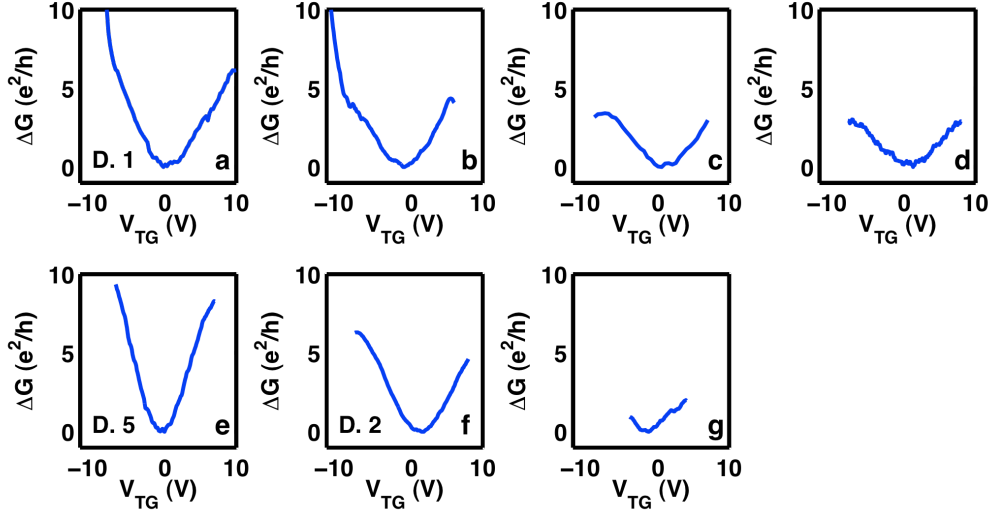

Figure S 3: The same $\Delta G(V_{TG})$ scans presented in Figure S1, with a contant slope Gsl = SL×$V_{TG}$ subtracted from each one.

To gain a quantitative estimate of the surface vs. bulk charging it is required to solve Poisson's equation within the depletion layer, accounting for all charge accumulated on the surface - including the topological states, metal-induced gap states, and all other surface defects. Precise accouting of all those surface effects requires a complex model which extends beyond the scope of this work. The slopes required to symmetrize each of the data sets presented in Figure S3 vary widely, and it is possible that this variability is related to differences in such surface details.

*Possible origin of the sharp decrease in resistance at negative gate voltage*
In some devices (e.g. D1 in Figure S3) a sharp increase in conductance appears at a negative gate voltage $V_{TG} \sim$ -8 V. This feature is difficult to investigate since it appears near the limit of the voltage accessible by the top gate. One possible origin is that at negative voltage the valence band is pulled above the Fermi energy, forming an inversion layer where electrical transport is carried by holes in the bulk (Figure S4).

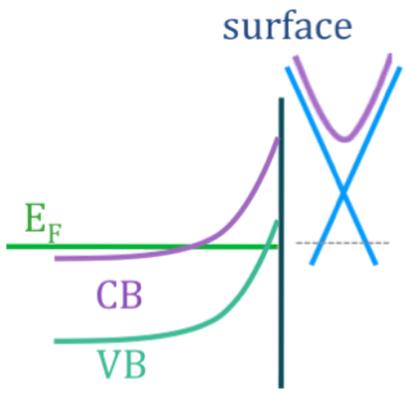

**Figure S4: Inversion layer formed at the top surface by the application of negative top-gate voltage.**